\documentclass[aps,prl,twocolumn,10pt,amsmath,amssymb,floats,floatfix,longbibliography,noeprint]{revtex4-1}
\usepackage{graphicx}
\usepackage{color}
\usepackage{float}
\usepackage{dcolumn}
\usepackage{amsmath}
\usepackage{bm}
\usepackage[utf8]{inputenc}
\usepackage{dsfont}

\begin{document}
\title{Nonreciprocal response theory of nonhermitian mechanical metamaterials:\\
response phase transition from the skin effect of zero modes}

\author{Henning Schomerus}
\affiliation{Department of Physics, Lancaster University, Lancaster LA1 4YB, United Kingdom}

\date{\today}

\begin{abstract}
Nonreciprocal nonhermitian systems provide an unconventional localization mechanism of topological zero modes
via the nonhermitian skin effect. While fundamental theoretical characterizations of this effect involve the biorthogonal system of right and left eigenmodes,
the recent demonstration of this effect for a zero mode in a robotic metamaterial
(Ghatak et al., arXiv:1907.11619)
is based on the direct experimental observation of the conventional right eigenvectors.
Here I show that such nonreciprocal mechanical metamaterials reveal their underlying biorthogonality in the directly observable response of the system to external excitation. Applied to the ground-breaking experiment, this nonreciprocal response theory predicts that the zero-mode skin effect goes along an extended phase where the system is highly sensitive to physical perturbations, leading to a diverging response in the limit of a large system.
\end{abstract}

\maketitle

Nonreciprocal nonhermitian mechanical metamaterials  are a recent innovation \cite{Brandenbourger.2019,Ghatak.2019}  that allow the experimental study of phenomena arising from the interplay of two principal notions of broken time-reversal symmetry in conservative and dissipative systems.
Nonreciprocal media break time-reversal symmetry via effective vector potentials, which translate into striking phenomena such as quantum-Hall like effects and optical isolation.
These effects are absent in the dissipative breaking of time-reversal symmetry by scalar gain and loss, resulting in nonhermitian physics where resonances acquire a finite life times.
Nonreciprocity and nonhermiticity are combined when one considers systems with directed gain or loss mechanisms, allowing them to sustain a finite net flux imbalance, as original introduced by Hatano and Nelson \cite{Hatano.1996}.
An only recently recognized striking consequence is the so-call nonhermitian skin effect, describing the possibility to localize states at the edge of the system when the imbalance is large enough, which requires to revisit the well-established bulk boundary correspondence known from hermitian systems  \cite{Xiong.2018,Yao.2018,Kunst.2018,Lieu.2018,McDonald.2018,Martinez.2018,Lee.2019,Brzezicki.2019}.

In a recent ground-breaking experiment, an analogous relocalization of a topological zero mode from one edge to the other has been realized in a robotic metamaterial \cite{Ghatak.2019}. The experimental observation of this zero-mode skin effect naturally maps out the spatial response of the system, which is linked to the right eigenmodes of the system. On the other hand, the theoretical understanding of the nonhermitian skin effect \cite{Kunst.2018} highlights the interplay of right and left eigenmodes, which is much more intricate than in reciprocal systems where both types of eigenmodes are simply related by time-reversal symmetry. In the nonreciprocal case, theory has to invoke biorthogonality relations that involve the complete set of eigenmodes of the system. This leaves the natural question of experimental signatures of this intricate interplay.

As I point out in this work, the left eigenmodes as well as the complete biorthogonal interplay both leave clear signatures that can be directly observed in experiments. These signatures are revealed when one develops the response theory for nonreciprocal media subjected to physical external excitation, which I here exemplify for a general class of systems compassing the robotic metamaterials.
The left eigenmodes characterize the strength of the response with respect to the location of the perturbation, while the right eigenmodes characterize the spatial distribution of the response itself, which has been in the focus of the experiments. Intriguingly, the nonhermitian skin effect of the zero mode then becomes linked with a phase transition, where the sensitivity of the system to low-frequency excitations diverges in the limit of a large system.
This extreme sensitivity, which occurs across the whole skin-effect phase and therefore is independent of any spectral singularities, is described by the formal analogue of the Petermann factor from quantum-limited noise theory \cite{Siegman.1989,Patra.2000,Berry.2003,Pick.2017}, and applies generally to a wide class of nonreciprocal nonhermitian media.

\begin{figure}[t]
\includegraphics[width=\columnwidth]{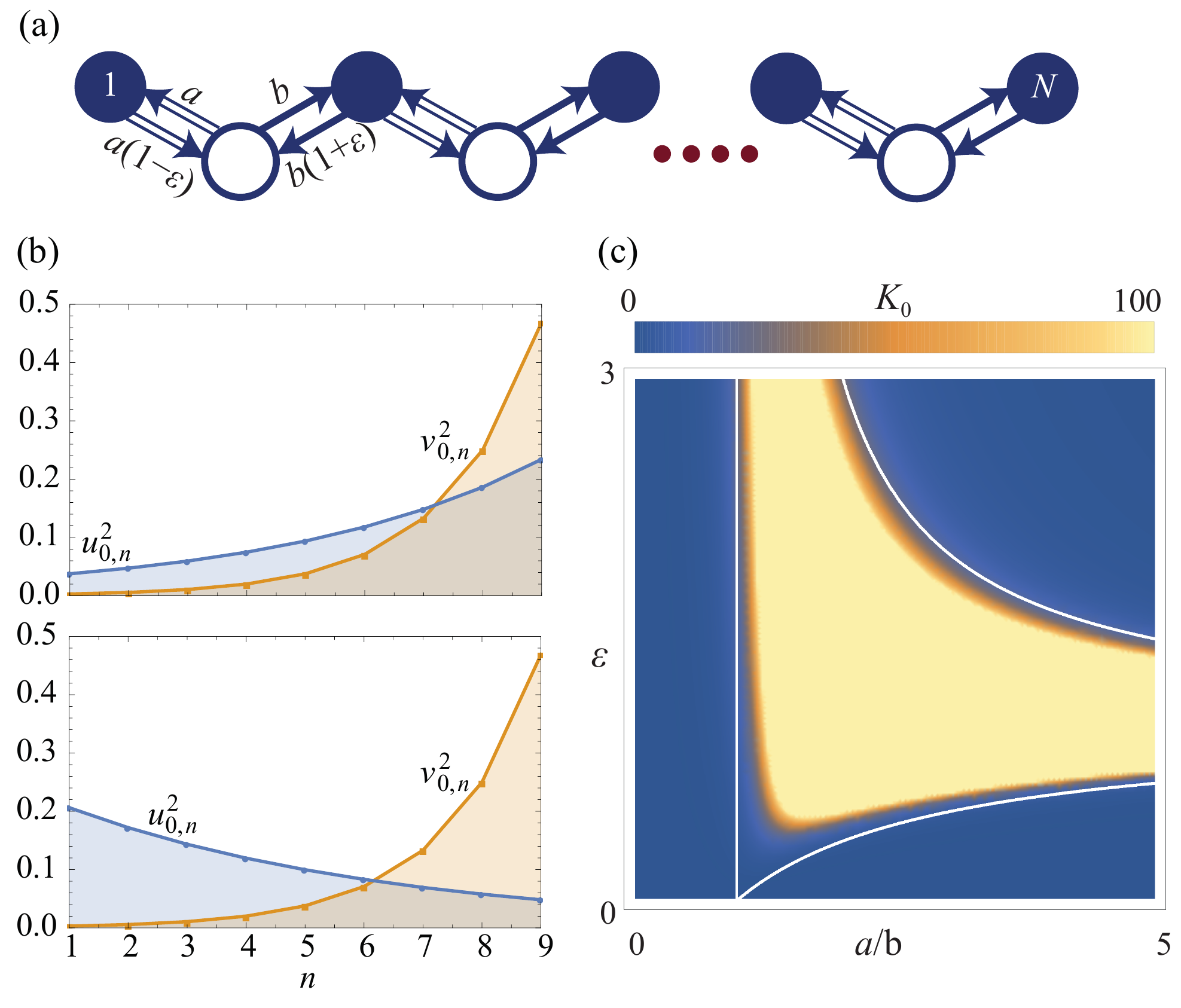}
  \caption{
(a) Nonreciprocal coupling configuration in a nonhermitian robotic metamaterial, using the unfolding
 \eqref{eq:unfolded} which maps the system onto a nonreciprocal Su-Schrieffer-Heeger chain with a topological zero mode.
 The full circles represent position degrees of freedom for rotors equipped with feedback, while the open circle relate to the resulting torque, which is not directly observed.
 This paper develops the nonreciprocal response theory for the rotor dynamics, which enters a phase of extreme sensitivity when the zero mode undergoes a relocalization from one edge to the other in analogy to the nonhermitian skin effect of bulk states, as shown in (b) for $a=1$, $b=0.73$, $N=9$, and $\varepsilon=0.1$ (top), $\varepsilon=0.2$ (bottom). (c) Resulting sensitivity phase diagram for finite $N=9$ in terms of the enhancement factor $K_0$, Eq.~\eqref{eq:k}. The white lines enclose the zero-mode skin-effect region.
 }
 \label{fig1}
\end{figure}

I develop the nonreciprocal response theory guided by
the robotic metamaterial in Ref.~\cite{Ghatak.2019}, in which $N$ coupled oscillators
$\frac{d^2\mathbf{x}}{dt^2}+M\mathbf{x}=0$
are equipped with a feedback force so that the dynamical matrix $M$ is asymmetric, $M\neq M^T$.
This realizes the directed couplings of a nonreciprocal nonhermitian system.
In the hermitian limit, the experimental system represents the Kane-Lubensky model of topological mechanics \cite{Kane.2014, Chen.2014,Huber.2016}, which heralds a topological zero mode due to the factorization of $M=QQ^T$ with an  $N\times (N-1)$-dimensional matrix $Q$, specifically chose to have elements
$Q_{nm}=-a\delta_{nm}+b\delta_{n-1,m}$.
The robotic metamaterial retains a modified factorization of the form
 $M=QR$, where the
 $(N-1)\times N$ matrix
$R$ with elements $R_{nm}=-a(1-\varepsilon)\delta_{nm}+b(1+\varepsilon)\delta_{n,m-1}$ differs from $Q^T$.
The parameter $\varepsilon$ quantifies the nonreciprocal couplings, and induces strong nonhermitian effects due to the two high-order exceptional points at $|\varepsilon|=1\equiv\varepsilon_{\mathrm{EP}}$, where all the bulk eigenmodes collapse onto a single eigenvector. Unfolding the system
as
\begin{equation}
H=\left(
\begin{array}{cc}
0 & Q
\\
R & 0
\end{array}
\right),\quad H^2=\mathrm{diag}(QR,RQ),
\label{eq:unfolded}
\end{equation}
the system maps onto the prototypical nonhermitian variant of a Su-Schrieffer-Heeger chain with nonreciprocal hoppings \cite{Yao.2018,Kunst.2018}, illustrated in Fig.~\ref{fig1}(a).

The dynamical modes of the system follow from the eigenvalue equations
\begin{align}
M\mathbf{u}_n=\Omega_n\mathbf{u}_n,\quad
\mathbf{v}_nM=\Omega_n\mathbf{v}_n,
\end{align}
with right and left eigenvectors $\mathbf{u}_n$ and $\mathbf{v}_n$.
Because of the ranks of matrices $Q$ and $R$ there is always a zero mode with $\Omega_0=0$, hence
$R\mathbf{u}_0
=0$, $\mathbf{v}_0 Q=0$, given by
\begin{align}
u_{0,n}=c_R \left(\frac{a(1-\varepsilon)}{b(1-\varepsilon)}\right)^n,
\quad v_{0,n}=c_L \left(\frac{a}{b}\right)^n
\label{eq:uv0}
\end{align}
with normalization constants $c_R$ and $c_L$.
The right eigenvector of the zero mode switches its localization position from one edge to the other at $\varepsilon_1=\frac{a-b}{a+b}$ and $\varepsilon_2=\frac{a+b}{a-b}=\varepsilon_1^{-1}$,
while the left eigenvector of the zero mode  always remains fixed in this design [Fig.~\ref{fig1}(b)]. As we will see, the mode can nonetheless directly be probed via the response of the system. In particular, the sensitivity of the system [Fig.~\ref{fig1}(c)] depends critically on the full biorthogonal interplay between the right and left modes, so that the invariability of the left mode is highly deceptive.

The main premise of this paper is the expectation that the response of such nonreciprocal nonhermitian systems to external perturbations should  be governed by a properly regularized Greens function
\begin{equation}
\hat G = (\omega^2\openone-M)^{-1}.
\label{eq:gf}
\end{equation}
Using the spectral decomposition
$M=U\,\hat\Omega^2 U^{-1}$, the Greens function includes the complete spectral information with eigenvalues $\Omega_n$ in the diagonal matrix $\hat\Omega$, as well as the full biorthogonal structure of eigenmodes with the corresponding right eigenvectors $\mathbf{u}_n$ as the columns of $U$, and the left eigenvectors $\mathbf{v}_n$ as the rows of $U^{-1}$.
In the presence of a zero mode, the Greens function has a double pole around $\omega=0$ complementing the simple poles at the bulk resonance frequencies---but this will not be the origin of the enhanced sensitivity, which instead arises from the emphasized role of mode biorthogonality in the zero-mode skin-effect phase.

To determine the  exact role of this Greens function, let us develop the detailed response theory of nonreciprocal nonhermitian mechanical media,
where for generality we do not invoke the factorization of $M$ nor assume the existence of a zero mode, hence also not restrict aspects such as dimensionality, order, coordination, or range of the couplings.
This also anticipates modified designs of nonreciprocal mechanical media  that either change the factorization so that left eigenvectors also change their localization position, or prevent factorization and remove the zero mode.
Furthermore, for generality we also allow modes to be complex.
Focussing on these general features of the system response then reveals the practical role of the right and left eigenvectors and leads to a characterization of the system in terms of its overall sensitivity.

Subject to quasi-harmonic external driving force
with a fixed force configuration $\mathbf{y}$, the response of the system is dictated by
\begin{equation}
\frac{d^2}{dt^2}\mathbf{x}+\gamma \frac{d}{dt}\mathbf{x} +M\mathbf{x}+\mathbf{y}\cos(\omega t)=0,
\end{equation}
where we include a velocity-dependent damping term of strength $\gamma$.
Considering some arbitrary initial conditions in the distant past, all finite-frequency components of the initial conditions will be damped out in the quasi-stationary response, but not those of the zero mode as the damping is velocity-dependent. We can remove this residual memory by also considering a finite width of the driving frequency, corresponding to a small imaginary part $\omega\pm i\eta$ in the advanced and retarded sectors of the description.

Using the spectral decomposition
$M=U\,\hat\Omega^2 U^{-1}$ mentioned above,
the quasi-stationary response is then given by
\begin{equation}
\mathbf{x}(t)=\mathrm{Re}\,\left[
U\,\left(\frac{e^{-i\hat\omega t}}{\hat\omega^2+i\hat\omega \gamma -\hat\Omega^2}\right)U^{-1}\right]\mathbf{y},
\end{equation}
where we set $\hat\omega=\omega\openone$.
This corresponds to the time-frequency Greens function
\begin{align}
G_{nm}(\omega;t)
&=\mathrm{Re}\,\sum_k
U_{nk}\left(\frac{e^{-i\omega t}}{\omega^2+i\omega\gamma-\Omega_k^2}\right)U^{-1}_{km}
,
\label{eq:gffinal}
\end{align}
giving the response at position $n$ for a unit-amplitude excitation at position $m$.
This expression naturally captures the spectrum as well as the right and left eigenvectors, where in particular the left eigenvectors describe how the response varies as one changes the location of the external drive.

\begin{figure}[t]
\includegraphics[width=\columnwidth]{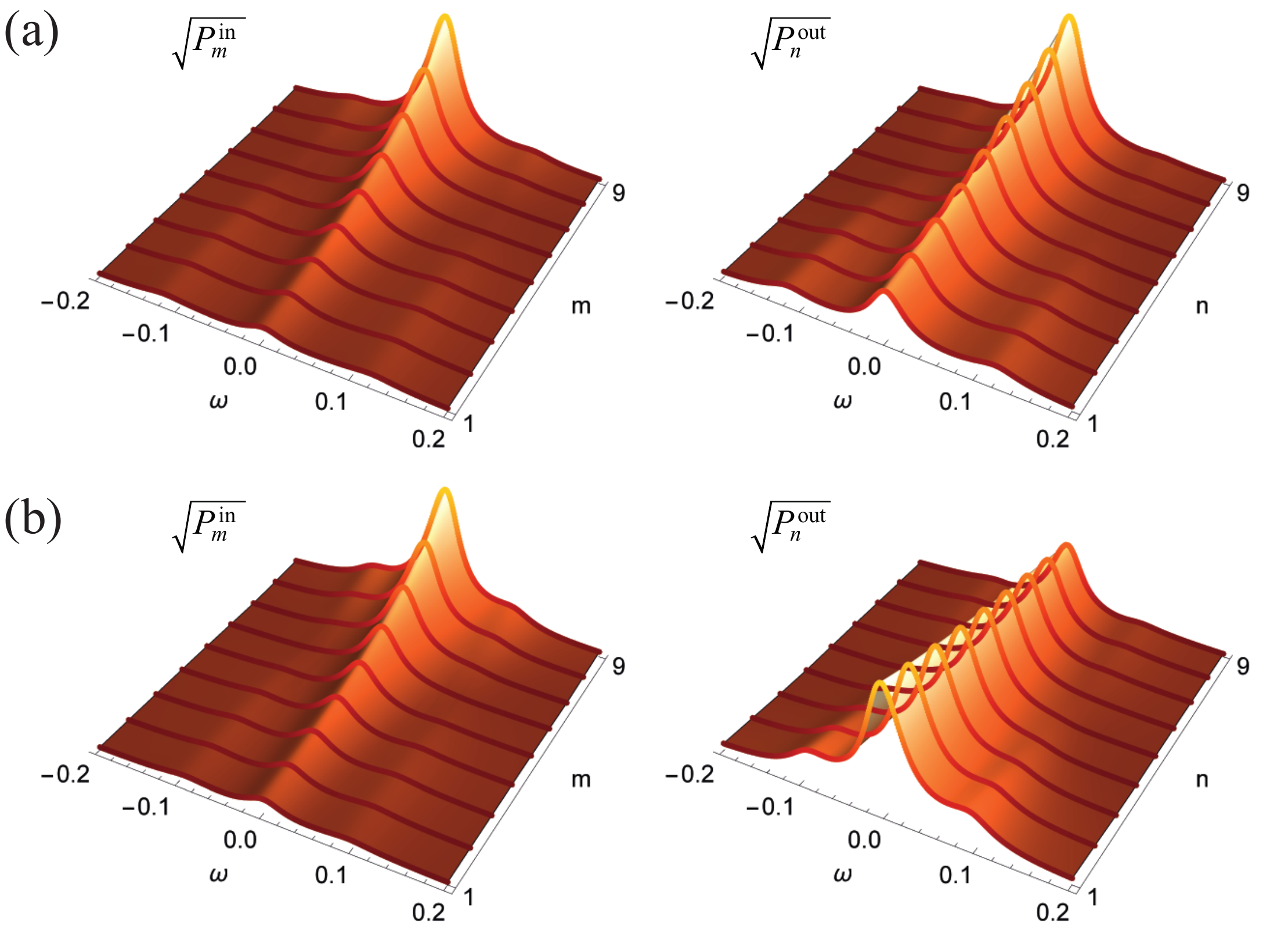}
  \caption{
  (a) Power amplitudes $\sqrt{P^{\mathrm{in}}_m(\omega)}$ (left) and $\sqrt{P^{\mathrm{out}}_n(\omega)}$  (right) [arb. units] for the nonreciprocal nonhermitian robotic metamaterial with $N=9$ components and parameters
$a=1$, $b=0.73$, $\varepsilon=0.1$, in the frequency range $|\omega|\leq 0.2$.
The guiding surfaces are interpolated in the discrete indices $n$ or $m$.
The central ridge corresponds to the zero mode, and maps out the corresponding left and right eigenvectors, respectively. The secondary ridges arise from the principal bulk mode, which resides at $\Omega_1=0.121$.
(b) Same for $\varepsilon=0.2$, beyond the critical value $\varepsilon_1=0.156$ at which the right eigenvector of the zero mode switches its localization centre. The neighbouring bulk mode resides at $\Omega_1=0.095$. The results are based on the regularization \eqref{eq:preg1} with $\eta=0.025$.
 }
 \label{fig2}
\end{figure}

These features of the system response can be further quantified using the
spatially resolved power spectrum for a unit-power excitation positioned at $m$ and detected at $n$,
\begin{align}
P_{nm}(\omega)
&=2\lim_{T\to \infty}\frac{1}{T}\int_0^T G^2_{nm}(\omega;t) dt
\\
&=
\sum_{kl}
\frac{U_{nk}U^{-1}_{km}}{\omega^2+i\omega\gamma-\Omega_k^2}\frac{(U_{nl}U^{-1}_{lm})^*}{\omega^2-i\omega\gamma-\Omega_l^{*2}}
,
\end{align}
where we can neglect the interference of the formally counter-rotating terms at $\omega\equiv 0$ as the power of the excitation has to be renormalized at this point as well (a unit-power excitation has amplitude $\propto \sqrt{2}\cos\omega t$ unless $\omega=0$, where the corresponding amplitude is unity).
We also define $P^{\mathrm{in}}_m(\omega)=\sum_n P_{nm}(\omega)$
for the collective system response to driving at position $m$, and the complementary
output spectrum $P^{\mathrm{out}}_n(\omega)=\sum_m P_{nm}(\omega)$
at position $n$ for
driving
across the whole system.
Close to resonance, where $\omega\approx \Omega_{\bar k}$ for a specific mode $\bar k$ \footnote{In the following expressions we assume for transparency real bare resonance frequencies, as these give rise to the most visible response features and naturally encompass the zero mode. From here the generalization to complex frequencies is straightforward, and only results in a further broadening, but retains the role of the right and left eigenvectors as well as the Petermann factor \eqref{eq:k}.},
\begin{align}
P^{\mathrm{in}}_m(\omega)& \approx
\frac{(U^\dagger U)_{\bar k \bar k}}{(\omega^2-\Omega_{\bar k}^2)^2+\omega^2\gamma^2}
|U^{-1}_{\bar k m}|^2
\end{align}
is then proportional to the intensity profile of the left eigenvector,
while
\begin{align}
P^{\mathrm{out}}_n(\omega)
& \approx
\frac{(U^{-1}U^{-\dagger})_{\bar k \bar k}}{(\omega^2-\Omega_{\bar k}^2)^2+\omega^2\gamma^2}
|U_{n \bar k}|^2
\end{align}
is proportional to the intensity profile of the right eigenvector.

Note that formally, the regularized results with a finite damping rate $\gamma$ are similar to the aforementioned implied frequency shift $\omega\to\omega \pm i\eta$ in the advanced and retarded Greens functions, which translates to the corresponding sectors of the power spectrum as
\begin{align}
\label{eq:preg1}
P_{nm}(\omega)=
\sum_{kl}
\frac{U_{nk}U^{-1}_{km}(U_{nl}U^{-1}_{lm})^*}{((\omega-i\eta)^2-\Omega_k^2)((\omega+i\eta)^2-\Omega_l^2)}.
\end{align}
However, the regularizations differ around $\omega=0$, where the physical velocity-dependent damping is ineffective.
In contrast, the regularization  \eqref{eq:preg1} corresponds to a finite width in the frequency  of the driving force itself.
In Fig.~\ref{fig2}, we illustrate the response of representative system configurations in terms of this frequency regularization. Using the values $a=1$ and $b=0.73$ from the experiment \cite{Ghatak.2019}, the zero-mode skin effect occurs at
$\varepsilon_1= 0.156$. The figure clearly shows how the system response is enhanced at opposite edges  for values on either side of this transition, following the relocalization of the right eigenmode. In contrast, the sensitivity of the system follows the invariable profile of the left eigenmode.

We now turn to the characterization of the system in terms of its overall sensitivity, which is captured by the total power spectrum $P^{\mathrm{tot}}(\omega)=\sum_{nm} P_{nm}(\omega)$.
Notably, the near resonance the overall response
\begin{align}
P^{\mathrm{tot}}(\omega)
 & \approx
\frac{K_{\bar k}}{(\omega^2-\Omega_{\bar k}^2)^2+\omega^2\gamma^2}
\end{align}
is then weighted by a factor
\begin{equation}
K_{\bar k}=(U^\dagger U)_{\bar k\bar k}(U^{-1}U^{-\dagger})_{{\bar k}{\bar k}}.
\label{eq:k}
\end{equation}
Mathematically, $K_{\bar k}\geq 1$ represents a condition number quantifying the mode nonorthogonality \cite{Chalker.1998},
while physically it signifies the ensuing enhanced sensitivity of the system to perturbations, in analogy to the Petermann factor from quantum-limited noise theory \cite{Siegman.1989,Patra.2000,Berry.2003}. The Petermann factor has been studied extensively for reciprocal nonhermitian systems \cite{Schomerus.2000,Fyodorov.2002,Yoo.2011,Pick.2017}, where $U^{-1}=U^T$ so that it can be calculated using only the right eigenvectors. In the present nonreciprocal case, however, we encounter a situation where this enhanced sensitivity involves the complete biorthogonal interplay of right and left eigenvectors.

\begin{figure}[t]
  \includegraphics[width=\columnwidth]{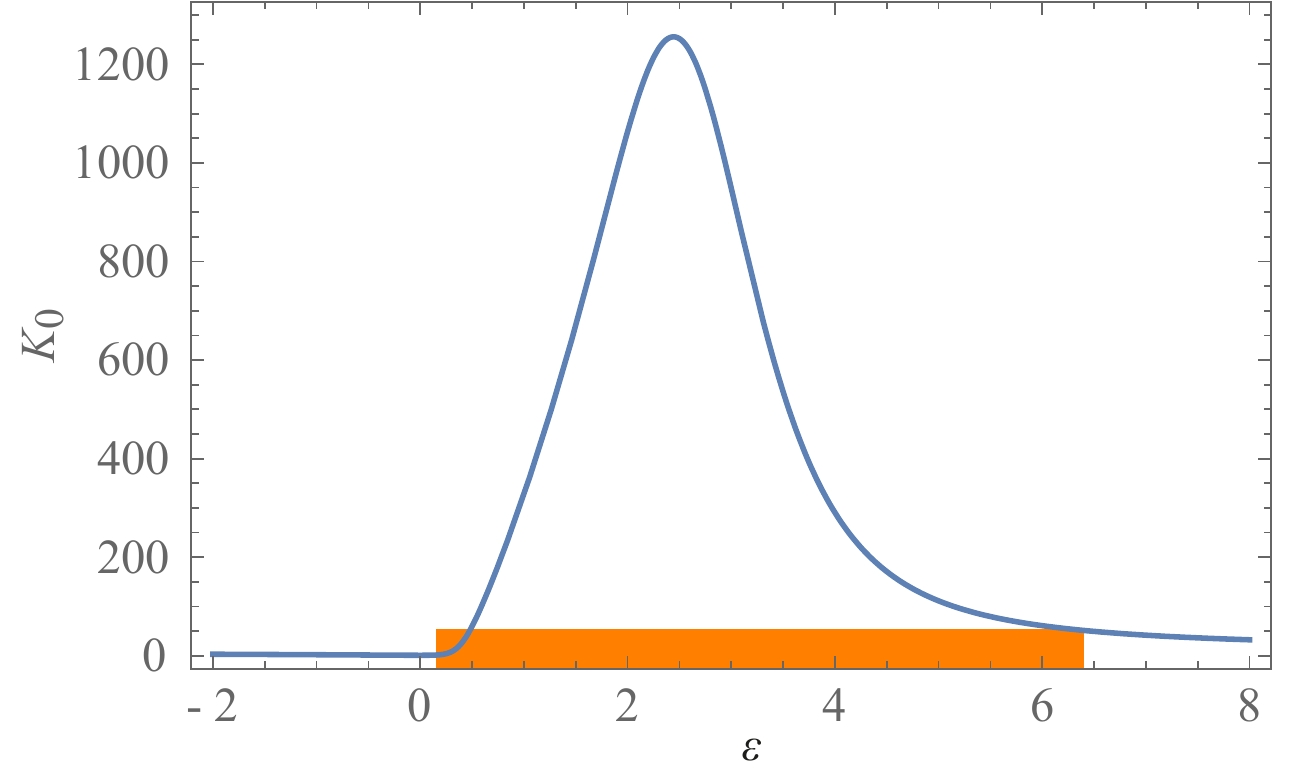}
  \caption{
  Petermann factor $K_0$  describing the enhanced sensitivity of the zero mode as a function of the nonhermiticity parameter $\varepsilon$, for a finite system with $N=9$, $a=1$ and $b=0.73$. The skin-effect phase of the zero mode is shaded on the horizontal axis.
  }\label{fig3}
 \end{figure}

Applying these results to the experimental setting, we first observe that the zero mode can have a strikingly large Petermann factor, despite its symmetry protection that distinguishes it from all other modes.
The known mode profiles
\eqref{eq:uv0} deliver the expression
\begin{align}
K_0=\frac{[\sum_{n=1}^N (a/b)^{2n}][\sum_{n=1}^N (a(1-\varepsilon)/b(1+\varepsilon))^{2n}]}{
[\sum_{n=1}^N (a^2(1-\varepsilon)/b^2(1+\varepsilon))^n]^2},
\end{align}
which is large in the zero-mode skin-effect phase $\varepsilon_1<\varepsilon<\varepsilon_2$.
The situation for finite $N=9$ is illustrated in Fig.~\ref{fig3}, which verifies that the Peterman factor rapidly increases when entering the phase with $\varepsilon>\varepsilon_1$, and again drops down leaving the phase at $\varepsilon>\varepsilon_2$.
Equipped with the skin effect,
the zero mode can therefore react  strongly to physical excitations, increasing its visibility in the experiments.

Notably, in the limit of a large system $N\to \infty$, the Petermann factor indeed remains finite only outside the skin-effect phase, where
\begin{equation}
K_0^{N\to\infty}\sim\frac{[a^2(1-\varepsilon)-b^2(1+\varepsilon)]^2}{
[a^2-b^2]
[a^2(1-\varepsilon)^2-b^2(1+\varepsilon)^2]},
\end{equation}
while $K_0$ diverges exponentially with increasing system size
in the range $\varepsilon_1<\varepsilon<\varepsilon_2$.
This fully reveals the advertised transition  to a phase of highly sensitive response, which is intimately tied to the nonhermitian skin effect of the zero mode. This correspondence emerges quickly across the whole parameter space already for moderate system sizes, as illustrated in Fig.~\ref{fig1}(c).

Note that in common nonhermitian settings, diverging sensitivities can occur via exceptional points \cite{Wiersig.2014}, i.e., spectral singularities at fine-tuned parameters in which eigenmodes collapse and the system becomes defective. The sensitive response phase identified here, however, is not tied to this mechanism. In particular, whilst being very large  for a finite system across the whole skin-effect phase, the sensitivity of the zero mode does not peak at the high-order bulk exceptional point $\varepsilon_{\mathrm{EP}}=1$, which strongly affects all the other modes according to $K_{n\neq 0}\sim [(\varepsilon_{\mathrm{EP}} - \varepsilon)/2)]^{1 - N}/N^2$ for $\varepsilon\to\varepsilon_{\mathrm{EP}}$
\footnote{With periodic boundary conditions, the bulk spectrum also displays exceptional points at $\varepsilon_1$ and $\varepsilon_2$, but these are not affecting the system with open boundary conditions \cite{Kunst.2018}.}. Furthermore, the transition occurs whilst the complete resonance spectrum remains real, hence does not resemble, e.g., the spontaneous breaking of nonhermitian symmetries as observed in PT-symmetric systems \cite{ElGanainy.2018}.

In summary, probing the dynamical response of  nonhermitian nonreciprocal metamaterials gives directly observable insights into the right and left eigenmodes, while the overall sensitivity of the system is governed by the full biorthogonal interplay between both sets of modes.
As we demonstrated for the example of a recently realized robotic metamaterial, the nonhermitian skin effect changing the localization position of a zero mode directly correlates with a  phase transition in which the medium becomes critically sensitive to perturbations.
These observations may be useful for sensing applications, which in contrast to earlier proposals invoking nonhermiticity \cite{Wiersig.2014,Chen.2017} would not rely on the closeness to an exceptional point and hence do not require the fine-tuning of parameters.

The general expressions of nonreciprocal nonhermitian response theory apply to linear systems with arbitrary dynamical matrix $M$, which can model linear systems of different dimensionality, coordination number, or range and disorder in the couplings, and  also serve as the starting point to derive continuum descriptions from microscopic models. Given the typical structures encountered for the right and left zero modes in the prototypical models studied so far, we expect this phase transition to be a general feature of nonreciprocal systems exhibiting a corresponding nonhermitian skin effect,
and possibly also extend to the skin effect of non-zero modes, as well as to
recent nonreciprocal and reciprocal variants of nonhermitian topoelectric circuits \cite{Helbig.2019,Hofmann.2019}.
From a more fundamental perspective,
the extreme sensitivity in the nonhermitian skin-effect phase outlines a practical limitation to stabilize novel nonhermitian phases, in analogy to what has transpired, e.g., for PT-symmetric systems
based on their sensitivity to quantum fluctuations \cite{Schomerus.2010,Scheel.2018}.

\begin{acknowledgments}
The author acknowledges funding by EPSRC via Grant No. EP/P010180/1 and
Programme Grant No. EP/N031776/1, as well as the hospitality of NORDITA where this work has been carried out.
\end{acknowledgments}


\begin{thebibliography}{31}%
\makeatletter
\providecommand \@ifxundefined [1]{%
 \@ifx{#1\undefined}
}%
\providecommand \@ifnum [1]{%
 \ifnum #1\expandafter \@firstoftwo
 \else \expandafter \@secondoftwo
 \fi
}%
\providecommand \@ifx [1]{%
 \ifx #1\expandafter \@firstoftwo
 \else \expandafter \@secondoftwo
 \fi
}%
\providecommand \natexlab [1]{#1}%
\providecommand \enquote  [1]{``#1''}%
\providecommand \bibnamefont  [1]{#1}%
\providecommand \bibfnamefont [1]{#1}%
\providecommand \citenamefont [1]{#1}%
\providecommand \href@noop [0]{\@secondoftwo}%
\providecommand \href [0]{\begingroup \@sanitize@url \@href}%
\providecommand \@href[1]{\@@startlink{#1}\@@href}%
\providecommand \@@href[1]{\endgroup#1\@@endlink}%
\providecommand \@sanitize@url [0]{\catcode `\\12\catcode `\$12\catcode
  `\&12\catcode `\#12\catcode `\^12\catcode `\_12\catcode `\%12\relax}%
\providecommand \@@startlink[1]{}%
\providecommand \@@endlink[0]{}%
\providecommand \url  [0]{\begingroup\@sanitize@url \@url }%
\providecommand \@url [1]{\endgroup\@href {#1}{\urlprefix }}%
\providecommand \urlprefix  [0]{URL }%
\providecommand \Eprint [0]{\href }%
\providecommand \doibase [0]{https://doi.org/}%
\providecommand \selectlanguage [0]{\@gobble}%
\providecommand \bibinfo  [0]{\@secondoftwo}%
\providecommand \bibfield  [0]{\@secondoftwo}%
\providecommand \translation [1]{[#1]}%
\providecommand \BibitemOpen [0]{}%
\providecommand \bibitemStop [0]{}%
\providecommand \bibitemNoStop [0]{.\EOS\space}%
\providecommand \EOS [0]{\spacefactor3000\relax}%
\providecommand \BibitemShut  [1]{\csname bibitem#1\endcsname}%
\let\auto@bib@innerbib\@empty
\bibitem [{\citenamefont {{Brandenbourger}}\ \emph {et~al.}(2019)\citenamefont
  {{Brandenbourger}}, \citenamefont {{Locsin}}, \citenamefont {{Lerner}},\ and\
  \citenamefont {{Coulais}}}]{Brandenbourger.2019}%
  \BibitemOpen
  \bibfield  {author} {\bibinfo {author} {\bibfnamefont {M.}~\bibnamefont
  {{Brandenbourger}}}, \bibinfo {author} {\bibfnamefont {X.}~\bibnamefont
  {{Locsin}}}, \bibinfo {author} {\bibfnamefont {E.}~\bibnamefont {{Lerner}}},\
  and\ \bibinfo {author} {\bibfnamefont {C.}~\bibnamefont {{Coulais}}},\
  }\bibfield  {title} {\bibinfo {title} {{Non-reciprocal robotic
  metamaterials}},\ }\href@noop {} {\bibfield  {journal} {\bibinfo  {journal}
  {arXiv e-prints arXiv:1903.03807}\ } (\bibinfo {year} {2019})}\BibitemShut
  {NoStop}%
\bibitem [{\citenamefont {{Ghatak}}\ \emph {et~al.}(2019)\citenamefont
  {{Ghatak}}, \citenamefont {{Brandenbourger}}, \citenamefont {{van Wezel}},\
  and\ \citenamefont {{Coulais}}}]{Ghatak.2019}%
  \BibitemOpen
  \bibfield  {author} {\bibinfo {author} {\bibfnamefont {A.}~\bibnamefont
  {{Ghatak}}}, \bibinfo {author} {\bibfnamefont {M.}~\bibnamefont
  {{Brandenbourger}}}, \bibinfo {author} {\bibfnamefont {J.}~\bibnamefont {{van
  Wezel}}},\ and\ \bibinfo {author} {\bibfnamefont {C.}~\bibnamefont
  {{Coulais}}},\ }\bibfield  {title} {\bibinfo {title} {Observation of
  non-hermitian topology and its bulk-edge correspondence},\ }\href@noop {}
  {\bibfield  {journal} {\bibinfo  {journal} {arXiv e-prints arXiv:1907.11619}\
  } (\bibinfo {year} {2019})}\BibitemShut {NoStop}%
\bibitem [{\citenamefont {Hatano}\ and\ \citenamefont
  {Nelson}(1996)}]{Hatano.1996}%
  \BibitemOpen
  \bibfield  {author} {\bibinfo {author} {\bibfnamefont {N.}~\bibnamefont
  {Hatano}}\ and\ \bibinfo {author} {\bibfnamefont {D.~R.}\ \bibnamefont
  {Nelson}},\ }\bibfield  {title} {\bibinfo {title} {Localization transitions
  in non-hermitian quantum mechanics},\ }\href
  {https://doi.org/10.1103/PhysRevLett.77.570} {\bibfield  {journal} {\bibinfo
  {journal} {Phys. Rev. Lett.}\ }\textbf {\bibinfo {volume} {77}},\ \bibinfo
  {pages} {570} (\bibinfo {year} {1996})}\BibitemShut {NoStop}%
\bibitem [{\citenamefont {Xiong}(2018)}]{Xiong.2018}%
  \BibitemOpen
  \bibfield  {author} {\bibinfo {author} {\bibfnamefont {Y.}~\bibnamefont
  {Xiong}},\ }\bibfield  {title} {\bibinfo {title} {Why does bulk boundary
  correspondence fail in some non-hermitian topological models},\ }\href
  {https://doi.org/10.1088/2399-6528/aab64a} {\bibfield  {journal} {\bibinfo
  {journal} {J. Phys. Commun.}\ }\textbf {\bibinfo {volume} {2}},\ \bibinfo
  {pages} {035043} (\bibinfo {year} {2018})}\BibitemShut {NoStop}%
\bibitem [{\citenamefont {Yao}\ and\ \citenamefont {Wang}(2018)}]{Yao.2018}%
  \BibitemOpen
  \bibfield  {author} {\bibinfo {author} {\bibfnamefont {S.}~\bibnamefont
  {Yao}}\ and\ \bibinfo {author} {\bibfnamefont {Z.}~\bibnamefont {Wang}},\
  }\bibfield  {title} {\bibinfo {title} {Edge states and topological invariants
  of non-hermitian systems},\ }\href
  {https://doi.org/10.1103/PhysRevLett.121.086803} {\bibfield  {journal}
  {\bibinfo  {journal} {Phys. Rev. Lett.}\ }\textbf {\bibinfo {volume} {121}},\
  \bibinfo {pages} {086803} (\bibinfo {year} {2018})}\BibitemShut {NoStop}%
\bibitem [{\citenamefont {Kunst}\ \emph {et~al.}(2018)\citenamefont {Kunst},
  \citenamefont {Edvardsson}, \citenamefont {Budich},\ and\ \citenamefont
  {Bergholtz}}]{Kunst.2018}%
  \BibitemOpen
  \bibfield  {author} {\bibinfo {author} {\bibfnamefont {F.~K.}\ \bibnamefont
  {Kunst}}, \bibinfo {author} {\bibfnamefont {E.}~\bibnamefont {Edvardsson}},
  \bibinfo {author} {\bibfnamefont {J.~C.}\ \bibnamefont {Budich}},\ and\
  \bibinfo {author} {\bibfnamefont {E.~J.}\ \bibnamefont {Bergholtz}},\
  }\bibfield  {title} {\bibinfo {title} {Biorthogonal bulk-boundary
  correspondence in non-hermitian systems},\ }\href
  {https://doi.org/10.1103/PhysRevLett.121.026808} {\bibfield  {journal}
  {\bibinfo  {journal} {Phys. Rev. Lett.}\ }\textbf {\bibinfo {volume} {121}},\
  \bibinfo {pages} {026808} (\bibinfo {year} {2018})}\BibitemShut {NoStop}%
\bibitem [{\citenamefont {Lieu}(2018)}]{Lieu.2018}%
  \BibitemOpen
  \bibfield  {author} {\bibinfo {author} {\bibfnamefont {S.}~\bibnamefont
  {Lieu}},\ }\bibfield  {title} {\bibinfo {title} {Topological phases in the
  non-hermitian su-schrieffer-heeger model},\ }\href
  {https://doi.org/10.1103/PhysRevB.97.045106} {\bibfield  {journal} {\bibinfo
  {journal} {Phys. Rev. B}\ }\textbf {\bibinfo {volume} {97}},\ \bibinfo
  {pages} {045106} (\bibinfo {year} {2018})}\BibitemShut {NoStop}%
\bibitem [{\citenamefont {McDonald}\ \emph {et~al.}(2018)\citenamefont
  {McDonald}, \citenamefont {Pereg-Barnea},\ and\ \citenamefont
  {Clerk}}]{McDonald.2018}%
  \BibitemOpen
  \bibfield  {author} {\bibinfo {author} {\bibfnamefont {A.}~\bibnamefont
  {McDonald}}, \bibinfo {author} {\bibfnamefont {T.}~\bibnamefont
  {Pereg-Barnea}},\ and\ \bibinfo {author} {\bibfnamefont {A.~A.}\ \bibnamefont
  {Clerk}},\ }\bibfield  {title} {\bibinfo {title} {Phase-dependent chiral
  transport and effective non-hermitian dynamics in a bosonic
  {Kitaev}-{Majorana} chain},\ }\href
  {https://doi.org/10.1103/PhysRevX.8.041031} {\bibfield  {journal} {\bibinfo
  {journal} {Phys. Rev. X}\ }\textbf {\bibinfo {volume} {8}},\ \bibinfo {pages}
  {041031} (\bibinfo {year} {2018})}\BibitemShut {NoStop}%
\bibitem [{\citenamefont {Martinez~Alvarez}\ \emph {et~al.}(2018)\citenamefont
  {Martinez~Alvarez}, \citenamefont {Barrios~Vargas},\ and\ \citenamefont
  {Foa~Torres}}]{Martinez.2018}%
  \BibitemOpen
  \bibfield  {author} {\bibinfo {author} {\bibfnamefont {V.~M.}\ \bibnamefont
  {Martinez~Alvarez}}, \bibinfo {author} {\bibfnamefont {J.~E.}\ \bibnamefont
  {Barrios~Vargas}},\ and\ \bibinfo {author} {\bibfnamefont {L.~E.~F.}\
  \bibnamefont {Foa~Torres}},\ }\bibfield  {title} {\bibinfo {title}
  {Non-hermitian robust edge states in one dimension: Anomalous localization
  and eigenspace condensation at exceptional points},\ }\href
  {https://doi.org/10.1103/PhysRevB.97.121401} {\bibfield  {journal} {\bibinfo
  {journal} {Phys. Rev. B}\ }\textbf {\bibinfo {volume} {97}},\ \bibinfo
  {pages} {121401(R)} (\bibinfo {year} {2018})}\BibitemShut {NoStop}%
\bibitem [{\citenamefont {Lee}\ and\ \citenamefont {Thomale}(2019)}]{Lee.2019}%
  \BibitemOpen
  \bibfield  {author} {\bibinfo {author} {\bibfnamefont {C.~H.}\ \bibnamefont
  {Lee}}\ and\ \bibinfo {author} {\bibfnamefont {R.}~\bibnamefont {Thomale}},\
  }\bibfield  {title} {\bibinfo {title} {Anatomy of skin modes and topology in
  non-hermitian systems},\ }\href {https://doi.org/10.1103/PhysRevB.99.201103}
  {\bibfield  {journal} {\bibinfo  {journal} {Phys. Rev. B}\ }\textbf {\bibinfo
  {volume} {99}},\ \bibinfo {pages} {201103(R)} (\bibinfo {year}
  {2019})}\BibitemShut {NoStop}%
\bibitem [{\citenamefont {{Brzezicki}}\ and\ \citenamefont
  {{Hyart}}(2019)}]{Brzezicki.2019}%
  \BibitemOpen
  \bibfield  {author} {\bibinfo {author} {\bibfnamefont {W.}~\bibnamefont
  {{Brzezicki}}}\ and\ \bibinfo {author} {\bibfnamefont {T.}~\bibnamefont
  {{Hyart}}},\ }\bibfield  {title} {\bibinfo {title} {{Hidden Chern number in
  one-dimensional non-Hermitian chiral-symmetric systems}},\ }\href@noop {}
  {\bibfield  {journal} {\bibinfo  {journal} {arXiv e-prints arXiv:1908.01553}\
  } (\bibinfo {year} {2019})}\BibitemShut {NoStop}%
\bibitem [{\citenamefont {Siegman}(1989)}]{Siegman.1989}%
  \BibitemOpen
  \bibfield  {author} {\bibinfo {author} {\bibfnamefont {A.~E.}\ \bibnamefont
  {Siegman}},\ }\bibfield  {title} {\bibinfo {title} {Excess spontaneous
  emission in non-hermitian optical systems. {I.} {Laser} amplifiers},\ }\href
  {https://doi.org/10.1103/PhysRevA.39.1253} {\bibfield  {journal} {\bibinfo
  {journal} {Phys. Rev. A}\ }\textbf {\bibinfo {volume} {39}},\ \bibinfo
  {pages} {1253} (\bibinfo {year} {1989})}\BibitemShut {NoStop}%
\bibitem [{\citenamefont {Patra}\ \emph {et~al.}(2000)\citenamefont {Patra},
  \citenamefont {Schomerus},\ and\ \citenamefont {Beenakker}}]{Patra.2000}%
  \BibitemOpen
  \bibfield  {author} {\bibinfo {author} {\bibfnamefont {M.}~\bibnamefont
  {Patra}}, \bibinfo {author} {\bibfnamefont {H.}~\bibnamefont {Schomerus}},\
  and\ \bibinfo {author} {\bibfnamefont {C.~W.~J.}\ \bibnamefont {Beenakker}},\
  }\bibfield  {title} {\bibinfo {title} {Quantum-limited linewidth of a chaotic
  laser cavity},\ }\href {https://doi.org/10.1103/PhysRevA.61.023810}
  {\bibfield  {journal} {\bibinfo  {journal} {Phys. Rev. A}\ }\textbf {\bibinfo
  {volume} {61}},\ \bibinfo {pages} {023810} (\bibinfo {year}
  {2000})}\BibitemShut {NoStop}%
\bibitem [{\citenamefont {Berry}(2003)}]{Berry.2003}%
  \BibitemOpen
  \bibfield  {author} {\bibinfo {author} {\bibfnamefont {M.~V.}\ \bibnamefont
  {Berry}},\ }\bibfield  {title} {\bibinfo {title} {Mode degeneracies and the
  petermann excess-noise factor for unstable lasers},\ }\href
  {https://doi.org/10.1080/09500340308234532} {\bibfield  {journal} {\bibinfo
  {journal} {J. Mod. Opt.}\ }\textbf {\bibinfo {volume} {50}},\ \bibinfo
  {pages} {63} (\bibinfo {year} {2003})}\BibitemShut {NoStop}%
\bibitem [{\citenamefont {Pick}\ \emph {et~al.}(2017)\citenamefont {Pick},
  \citenamefont {Zhen}, \citenamefont {Miller}, \citenamefont {Hsu},
  \citenamefont {Hernandez}, \citenamefont {Rodriguez}, \citenamefont
  {Solja\v{c}i\'{c}},\ and\ \citenamefont {Johnson}}]{Pick.2017}%
  \BibitemOpen
  \bibfield  {author} {\bibinfo {author} {\bibfnamefont {A.}~\bibnamefont
  {Pick}}, \bibinfo {author} {\bibfnamefont {B.}~\bibnamefont {Zhen}}, \bibinfo
  {author} {\bibfnamefont {O.~D.}\ \bibnamefont {Miller}}, \bibinfo {author}
  {\bibfnamefont {C.~W.}\ \bibnamefont {Hsu}}, \bibinfo {author} {\bibfnamefont
  {F.}~\bibnamefont {Hernandez}}, \bibinfo {author} {\bibfnamefont {A.~W.}\
  \bibnamefont {Rodriguez}}, \bibinfo {author} {\bibfnamefont {M.}~\bibnamefont
  {Solja\v{c}i\'{c}}},\ and\ \bibinfo {author} {\bibfnamefont {S.~G.}\
  \bibnamefont {Johnson}},\ }\bibfield  {title} {\bibinfo {title} {General
  theory of spontaneous emission near exceptional points},\ }\href
  {https://doi.org/10.1364/OE.25.012325} {\bibfield  {journal} {\bibinfo
  {journal} {Opt. Express}\ }\textbf {\bibinfo {volume} {25}},\ \bibinfo
  {pages} {12325} (\bibinfo {year} {2017})}\BibitemShut {NoStop}%
\bibitem [{\citenamefont {Kane}\ and\ \citenamefont
  {Lubensky}(2014)}]{Kane.2014}%
  \BibitemOpen
  \bibfield  {author} {\bibinfo {author} {\bibfnamefont {C.~L.}\ \bibnamefont
  {Kane}}\ and\ \bibinfo {author} {\bibfnamefont {T.~C.}\ \bibnamefont
  {Lubensky}},\ }\bibfield  {title} {\bibinfo {title} {Topological boundary
  modes in isostatic lattices},\ }\href {https://doi.org/10.1038/nphys2835}
  {\bibfield  {journal} {\bibinfo  {journal} {Nature Phys.}\ }\textbf {\bibinfo
  {volume} {10}},\ \bibinfo {pages} {39} (\bibinfo {year} {2014})}\BibitemShut
  {NoStop}%
\bibitem [{\citenamefont {Chen}\ \emph {et~al.}(2014)\citenamefont {Chen},
  \citenamefont {Upadhyaya},\ and\ \citenamefont {Vitelli}}]{Chen.2014}%
  \BibitemOpen
  \bibfield  {author} {\bibinfo {author} {\bibfnamefont {B.~G.}\ \bibnamefont
  {Chen}}, \bibinfo {author} {\bibfnamefont {N.}~\bibnamefont {Upadhyaya}},\
  and\ \bibinfo {author} {\bibfnamefont {V.}~\bibnamefont {Vitelli}},\
  }\bibfield  {title} {\bibinfo {title} {Nonlinear conduction via solitons in a
  topological mechanical insulator},\ }\href
  {https://doi.org/10.1073/pnas.1405969111} {\bibfield  {journal} {\bibinfo
  {journal} {PNAS}\ }\textbf {\bibinfo {volume} {111}},\ \bibinfo {pages}
  {13004} (\bibinfo {year} {2014})}\BibitemShut {NoStop}%
\bibitem [{\citenamefont {Huber}(2016)}]{Huber.2016}%
  \BibitemOpen
  \bibfield  {author} {\bibinfo {author} {\bibfnamefont {S.~D.}\ \bibnamefont
  {Huber}},\ }\bibfield  {title} {\bibinfo {title} {Topological mechanics},\
  }\href {https://doi.org/10.1038/nphys3801} {\bibfield  {journal} {\bibinfo
  {journal} {Nature Phys.}\ }\textbf {\bibinfo {volume} {12}},\ \bibinfo
  {pages} {621 EP } (\bibinfo {year} {2016})}\BibitemShut {NoStop}%
\bibitem [{Note1()}]{Note1}%
  \BibitemOpen
  \bibinfo {note} {In the following expressions we assume for transparency real
  bare resonance frequencies, as these give rise to the most visible response
  features and naturally encompass the zero mode. From here the generalization
  to complex frequencies is straightforward, and only results in a further
  broadening, but retains the role of the right and left eigenvectors as well
  as the Petermann factor \protect \textup {\hbox {\mathsurround \z@ \protect
  \normalfont (\ignorespaces \ref {eq:k}\unskip \@@italiccorr )}}.}\BibitemShut
  {Stop}%
\bibitem [{\citenamefont {Chalker}\ and\ \citenamefont
  {Mehlig}(1998)}]{Chalker.1998}%
  \BibitemOpen
  \bibfield  {author} {\bibinfo {author} {\bibfnamefont {J.~T.}\ \bibnamefont
  {Chalker}}\ and\ \bibinfo {author} {\bibfnamefont {B.}~\bibnamefont
  {Mehlig}},\ }\bibfield  {title} {\bibinfo {title} {Eigenvector statistics in
  non-hermitian random matrix ensembles},\ }\href
  {https://doi.org/10.1103/PhysRevLett.81.3367} {\bibfield  {journal} {\bibinfo
   {journal} {Phys. Rev. Lett.}\ }\textbf {\bibinfo {volume} {81}},\ \bibinfo
  {pages} {3367} (\bibinfo {year} {1998})}\BibitemShut {NoStop}%
\bibitem [{\citenamefont {Schomerus}\ \emph {et~al.}(2000)\citenamefont
  {Schomerus}, \citenamefont {Frahm}, \citenamefont {Patra},\ and\
  \citenamefont {Beenakker}}]{Schomerus.2000}%
  \BibitemOpen
  \bibfield  {author} {\bibinfo {author} {\bibfnamefont {H.}~\bibnamefont
  {Schomerus}}, \bibinfo {author} {\bibfnamefont {K.}~\bibnamefont {Frahm}},
  \bibinfo {author} {\bibfnamefont {M.}~\bibnamefont {Patra}},\ and\ \bibinfo
  {author} {\bibfnamefont {C.}~\bibnamefont {Beenakker}},\ }\bibfield  {title}
  {\bibinfo {title} {Quantum limit of the laser line width in chaotic cavities
  and statistics of residues of scattering matrix poles},\ }\href
  {https://doi.org/{10.1016/S0378-4371(99)00602-0}} {\bibfield  {journal}
  {\bibinfo  {journal} {Physica A}\ }\textbf {\bibinfo {volume} {278}},\
  \bibinfo {pages} {469} (\bibinfo {year} {2000})}\BibitemShut {NoStop}%
\bibitem [{\citenamefont {Fyodorov}\ and\ \citenamefont
  {Mehlig}(2002)}]{Fyodorov.2002}%
  \BibitemOpen
  \bibfield  {author} {\bibinfo {author} {\bibfnamefont {Y.~V.}\ \bibnamefont
  {Fyodorov}}\ and\ \bibinfo {author} {\bibfnamefont {B.}~\bibnamefont
  {Mehlig}},\ }\bibfield  {title} {\bibinfo {title} {Statistics of resonances
  and nonorthogonal eigenfunctions in a model for single-channel chaotic
  scattering},\ }\href {https://doi.org/10.1103/PhysRevE.66.045202} {\bibfield
  {journal} {\bibinfo  {journal} {Phys. Rev. E}\ }\textbf {\bibinfo {volume}
  {66}},\ \bibinfo {pages} {045202(R)} (\bibinfo {year} {2002})}\BibitemShut
  {NoStop}%
\bibitem [{\citenamefont {Yoo}\ \emph {et~al.}(2011)\citenamefont {Yoo},
  \citenamefont {Sim},\ and\ \citenamefont {Schomerus}}]{Yoo.2011}%
  \BibitemOpen
  \bibfield  {author} {\bibinfo {author} {\bibfnamefont {G.}~\bibnamefont
  {Yoo}}, \bibinfo {author} {\bibfnamefont {H.-S.}\ \bibnamefont {Sim}},\ and\
  \bibinfo {author} {\bibfnamefont {H.}~\bibnamefont {Schomerus}},\ }\bibfield
  {title} {\bibinfo {title} {Quantum noise and mode nonorthogonality in
  non-hermitian {PT}-symmetric optical resonators},\ }\href
  {https://doi.org/10.1103/PhysRevA.84.063833} {\bibfield  {journal} {\bibinfo
  {journal} {Phys. Rev. A}\ }\textbf {\bibinfo {volume} {84}},\ \bibinfo
  {pages} {063833} (\bibinfo {year} {2011})}\BibitemShut {NoStop}%
\bibitem [{\citenamefont {Wiersig}(2014)}]{Wiersig.2014}%
  \BibitemOpen
  \bibfield  {author} {\bibinfo {author} {\bibfnamefont {J.}~\bibnamefont
  {Wiersig}},\ }\bibfield  {title} {\bibinfo {title} {Enhancing the sensitivity
  of frequency and energy splitting detection by using exceptional points:
  Application to microcavity sensors for single-particle detection},\ }\href
  {https://doi.org/10.1103/PhysRevLett.112.203901} {\bibfield  {journal}
  {\bibinfo  {journal} {Phys. Rev. Lett.}\ }\textbf {\bibinfo {volume} {112}},\
  \bibinfo {pages} {203901} (\bibinfo {year} {2014})}\BibitemShut {NoStop}%
\bibitem [{Note2()}]{Note2}%
  \BibitemOpen
  \bibinfo {note} {With periodic boundary conditions, the bulk spectrum also
  displays exceptional points at $\varepsilon _1$ and $\varepsilon _2$, but
  these are not affecting the system with open boundary conditions \cite
  {Kunst.2018}.}\BibitemShut {Stop}%
\bibitem [{\citenamefont {El-Ganainy}\ \emph {et~al.}(2018)\citenamefont
  {El-Ganainy}, \citenamefont {Makris}, \citenamefont {Khajavikhan},
  \citenamefont {Musslimani}, \citenamefont {Rotter},\ and\ \citenamefont
  {Christodoulides}}]{ElGanainy.2018}%
  \BibitemOpen
  \bibfield  {author} {\bibinfo {author} {\bibfnamefont {R.}~\bibnamefont
  {El-Ganainy}}, \bibinfo {author} {\bibfnamefont {K.~G.}\ \bibnamefont
  {Makris}}, \bibinfo {author} {\bibfnamefont {M.}~\bibnamefont {Khajavikhan}},
  \bibinfo {author} {\bibfnamefont {Z.~H.}\ \bibnamefont {Musslimani}},
  \bibinfo {author} {\bibfnamefont {S.}~\bibnamefont {Rotter}},\ and\ \bibinfo
  {author} {\bibfnamefont {D.~N.}\ \bibnamefont {Christodoulides}},\ }\bibfield
   {title} {\bibinfo {title} {Non-hermitian physics and {PT} symmetry},\ }\href
  {https://doi.org/10.1038/nphys4323} {\bibfield  {journal} {\bibinfo
  {journal} {Nature Phys.}\ }\textbf {\bibinfo {volume} {14}},\ \bibinfo
  {pages} {11} (\bibinfo {year} {2018})}\BibitemShut {NoStop}%
\bibitem [{\citenamefont {Chen}\ \emph {et~al.}(2017)\citenamefont {Chen},
  \citenamefont {{Kaya {\"O}zdemir}}, \citenamefont {Zhao}, \citenamefont
  {Wiersig},\ and\ \citenamefont {Yang}}]{Chen.2017}%
  \BibitemOpen
  \bibfield  {author} {\bibinfo {author} {\bibfnamefont {W.}~\bibnamefont
  {Chen}}, \bibinfo {author} {\bibfnamefont {{\c{S}}.}~\bibnamefont {{Kaya
  {\"O}zdemir}}}, \bibinfo {author} {\bibfnamefont {G.}~\bibnamefont {Zhao}},
  \bibinfo {author} {\bibfnamefont {J.}~\bibnamefont {Wiersig}},\ and\ \bibinfo
  {author} {\bibfnamefont {L.}~\bibnamefont {Yang}},\ }\bibfield  {title}
  {\bibinfo {title} {Exceptional points enhance sensing in an optical
  microcavity},\ }\href {https://doi.org/10.1038/nature23281} {\bibfield
  {journal} {\bibinfo  {journal} {Nature}\ }\textbf {\bibinfo {volume} {548}},\
  \bibinfo {pages} {192} (\bibinfo {year} {2017})}\BibitemShut {NoStop}%
\bibitem [{\citenamefont {{Helbig}}\ \emph {et~al.}(2019)\citenamefont
  {{Helbig}}, \citenamefont {{Hofmann}}, \citenamefont {{Imhof}}, \citenamefont
  {{Abdelghany}}, \citenamefont {{Kiessling}}, \citenamefont {{Molenkamp}},
  \citenamefont {{Lee}}, \citenamefont {{Szameit}}, \citenamefont {{Greiter}},\
  and\ \citenamefont {{Thomale}}}]{Helbig.2019}%
  \BibitemOpen
  \bibfield  {author} {\bibinfo {author} {\bibfnamefont {T.}~\bibnamefont
  {{Helbig}}}, \bibinfo {author} {\bibfnamefont {T.}~\bibnamefont {{Hofmann}}},
  \bibinfo {author} {\bibfnamefont {S.}~\bibnamefont {{Imhof}}}, \bibinfo
  {author} {\bibfnamefont {M.}~\bibnamefont {{Abdelghany}}}, \bibinfo {author}
  {\bibfnamefont {T.}~\bibnamefont {{Kiessling}}}, \bibinfo {author}
  {\bibfnamefont {L.~W.}\ \bibnamefont {{Molenkamp}}}, \bibinfo {author}
  {\bibfnamefont {C.~H.}\ \bibnamefont {{Lee}}}, \bibinfo {author}
  {\bibfnamefont {A.}~\bibnamefont {{Szameit}}}, \bibinfo {author}
  {\bibfnamefont {M.}~\bibnamefont {{Greiter}}},\ and\ \bibinfo {author}
  {\bibfnamefont {R.}~\bibnamefont {{Thomale}}},\ }\bibfield  {title} {\bibinfo
  {title} {{Observation of bulk boundary correspondence breakdown in
  topolectrical circuits}},\ }\href@noop {} {\bibfield  {journal} {\bibinfo
  {journal} {arXiv e-prints arXiv:1907.11562}\ } (\bibinfo {year}
  {2019})}\BibitemShut {NoStop}%
\bibitem [{\citenamefont {{Hofmann}}\ \emph {et~al.}(2019)\citenamefont
  {{Hofmann}}, \citenamefont {{Helbig}}, \citenamefont {{Schindler}},
  \citenamefont {{Salgo}}, \citenamefont {{Brzezi{\'n}ska}}, \citenamefont
  {{Greiter}}, \citenamefont {{Kiessling}}, \citenamefont {{Wolf}},
  \citenamefont {{Vollhardt}}, \citenamefont {{Kaba{\v{s}}i}}, \citenamefont
  {{Lee}}, \citenamefont {{Bilu{\v{s}}i{\'c}}}, \citenamefont {{Thomale}},\
  and\ \citenamefont {{Neupert}}}]{Hofmann.2019}%
  \BibitemOpen
  \bibfield  {author} {\bibinfo {author} {\bibfnamefont {T.}~\bibnamefont
  {{Hofmann}}}, \bibinfo {author} {\bibfnamefont {T.}~\bibnamefont {{Helbig}}},
  \bibinfo {author} {\bibfnamefont {F.}~\bibnamefont {{Schindler}}}, \bibinfo
  {author} {\bibfnamefont {N.}~\bibnamefont {{Salgo}}}, \bibinfo {author}
  {\bibfnamefont {M.}~\bibnamefont {{Brzezi{\'n}ska}}}, \bibinfo {author}
  {\bibfnamefont {M.}~\bibnamefont {{Greiter}}}, \bibinfo {author}
  {\bibfnamefont {T.}~\bibnamefont {{Kiessling}}}, \bibinfo {author}
  {\bibfnamefont {D.}~\bibnamefont {{Wolf}}}, \bibinfo {author} {\bibfnamefont
  {A.}~\bibnamefont {{Vollhardt}}}, \bibinfo {author} {\bibfnamefont
  {A.}~\bibnamefont {{Kaba{\v{s}}i}}}, \bibinfo {author} {\bibfnamefont
  {C.~H.}\ \bibnamefont {{Lee}}}, \bibinfo {author} {\bibfnamefont
  {A.}~\bibnamefont {{Bilu{\v{s}}i{\'c}}}}, \bibinfo {author} {\bibfnamefont
  {R.}~\bibnamefont {{Thomale}}},\ and\ \bibinfo {author} {\bibfnamefont
  {T.}~\bibnamefont {{Neupert}}},\ }\bibfield  {title} {\bibinfo {title}
  {{Reciprocal skin effect and its realization in a topolectrical circuit}},\
  }\href@noop {} {\bibfield  {journal} {\bibinfo  {journal} {arXiv e-prints
  arXiv:1908.02759}\ } (\bibinfo {year} {2019})}\BibitemShut {NoStop}%
\bibitem [{\citenamefont {Schomerus}(2010)}]{Schomerus.2010}%
  \BibitemOpen
  \bibfield  {author} {\bibinfo {author} {\bibfnamefont {H.}~\bibnamefont
  {Schomerus}},\ }\bibfield  {title} {\bibinfo {title} {Quantum noise and
  self-sustained radiation of {PT}-symmetric systems},\ }\href
  {https://doi.org/{10.1103/PhysRevLett.104.233601}} {\bibfield  {journal}
  {\bibinfo  {journal} {{Phys. Rev. Lett.}}\ }\textbf {\bibinfo {volume}
  {104}},\ \bibinfo {pages} {233601} (\bibinfo {year} {2010})}\BibitemShut
  {NoStop}%
\bibitem [{\citenamefont {Scheel}\ and\ \citenamefont
  {Szameit}(2018)}]{Scheel.2018}%
  \BibitemOpen
  \bibfield  {author} {\bibinfo {author} {\bibfnamefont {S.}~\bibnamefont
  {Scheel}}\ and\ \bibinfo {author} {\bibfnamefont {A.}~\bibnamefont
  {Szameit}},\ }\bibfield  {title} {\bibinfo {title} {{PT}-symmetric photonic
  quantum systems with gain and loss do not exist},\ }\href
  {https://doi.org/10.1209/0295-5075/122/34001} {\bibfield  {journal} {\bibinfo
   {journal} {{EPL}}\ }\textbf {\bibinfo {volume} {122}},\ \bibinfo {pages}
  {34001} (\bibinfo {year} {2018})}\BibitemShut {NoStop}%
\end{thebibliography}

%

\end{document}